# Flat Thomas-Fermi artificial atoms

Yu. N. Ovchinnikov,[1] Avik Halder,[2] and Vitaly V. Kresin[2]

[1] *L. D. Landau Institute for Theoretical Physics, Russian Academy of Sciences, 142432, Chernogolovka, Moscow Region, Russia*

*and Max Planck Institute for the Physics of Complex Systems, D-01187, Dresden, Germany*

[2] *Department of Physics and Astronomy, University of Southern California, Los Angeles, California 90089-0484, USA*

## Abstract

We consider two-dimensional (2D) "artificial atoms" confined by an axially symmetric potential $V(\rho)$. Such configurations arise in circular quantum dots and other systems effectively restricted to a 2D layer. Using the semiclassical method, we present the first fully self-consistent and analytic solution yielding equations describing the density distribution, energy, and other quantities for any form of $V(\rho)$ and an arbitrary number of confined particles. An essential and nontrivial aspect of the problem is that the 2D density of states must be properly combined with 3D electrostatics. The solution turns out to have a universal form, with scaling parameters $\rho^2/R^2$ and $R/a_B^*$ ($R$ is the dot radius and $a_B^*$ is the effective Bohr radius).

**Introduction.** - The semiclassical Thomas-Fermi (TF) treatment of many-electron atoms [1,2] is recognized as an amalgam of physical insight and theoretical elegance. As is well-known and extensively reviewed (see, e.g., Refs. [3-10]), it has been widely applied and augmented. In this paper we focus on the two-dimensional (2D) analogue to the TF atom, such as found, e.g., in circularly symmetric semiconductor quantum dots (QD, frequently referred to as "artificial atoms" [11,12]). The motivation for this work is as follows. First of all, we obtain fully analytical solutions for the electron density distribution, energy, and other quantum dot parameters. Since it is known (see, e.g, Refs. [13-15] and references therein) that the semiclassical treatment of quantum dots often yields quantitatively accurate results (cf. Fig. 3 below), the recognized utility of analytical expressions is that they signpost the dependence of dot parameters on materials properties, electron number, and confinement potential strengths over a wide range of variations. Secondly, the mathematical structure of the 2D solution is quite interesting and provides a conceptually valuable complement to the textbook 3D solutions. And thirdly, the solution identifies nontrivial edge singularities in the electron density distribution; it is important to take them into account in constructing numerical solutions and variational functions for density-functional analyses of two-dimensional nanosystems.

The essence of the TF statistical method is to relate the electron number density at every point, $n(\vec{r})$, in two ways to the self-consistent electrostatic potential $\varphi(\vec{r})$ generated by these same electrons. On one hand, $\varphi$ satisfies the Poisson equation. On the other hand, the maximum kinetic energy of the electron gas, treated semiclassically, cannot exceed the local depth of the potential well. Thus the standard form of the TF differential equation derives from $p_F^2(\vec{r})/2m_e + e\varphi(\vec{r}) + V(\vec{r}) = \mu$ and $\nabla^2\varphi(\vec{r}) = -4\pi e n(\vec{r})$, where $m_e$ and $e<0$ are the electron mass and charge, $p_F$ is the local value of the Fermi momentum, $V$ is the externally applied potential, and $\mu$ is the chemical potential of the electron system ($\mu=0$ for a neutral isolated atom, but not in general).

In three dimensions the kinetic energy term is proportional to $n^{2/3}$ and the equation is nonlinear.[*] Consequently, it is very interesting to consider the 2D case realized, as mentioned above, in semiconductor QD islands. Here the electrons are restricted to a thin layer, with their transverse motion quantized but in-plane motion treated as that of an electron cloud confined in the radial direction by a potential $V(\vec{\rho})$ created by external electrostatic gates ($\vec{\rho}$ is the 2D radius vector). Below, we focus on the case of circular (axially symmetric) QD with $V(\vec{\rho}) = V(\rho)$; ellipsoidal shapes will be considered elsewhere. Also, for brevity we assume strong size quantization with only the lowest transverse level occupied.

The point of interest is that, as is well-known, the density of translational states in two dimensions is a constant and the kinetic energy at the Fermi level is then simply proportional to the density: $p_F^2(\vec{\rho}) = 2\pi\hbar^2 n(\vec{\rho})$. This raises the appealing prospect of a linear TF equation.

However, as happens all too often, there is a complication. While the electron cloud density and the confining potential are functions of the 2D radial coordinate, the electrostatic potential satisfies the Poisson equation in three-dimensional space. Thus what we actually face is a *linear*, but *three-dimensional* equation, essentially relating $\varphi(\vec{r})$ with $n(\vec{\rho})\delta(z)$.

---

[*] It can be approximately linearized in certain cases, such as when $n(\vec{r})$ deviates weakly from a uniform distribution (for example, impurity screening in metals [16] and the interior of metal nanoclusters [9]).

A number of papers simply replace the 3D Laplacian in the Poisson equation by a 2D one. This makes the differential equation elementary to solve, but of course corresponds not to a "pancake" of point electrons, but to a pool of infinitely long parallel line charges each generating a logarithmic, rather than a point-Coulomb, potential. Thus it is either a misrepresentation of the actual QD problem [17-20] or simply an interesting but abstract exercise [21-25].

In Ref. [26] the TF solution for QD electrons was pursued properly in 3D. The in-plane density $n(\rho)$ and potential $\varphi(\rho,z=0)$ were sought in the form of a 2D radial Fourier-Bessel series for $0<\rho<R$ ($R$ is the radius of the confined electron cloud). However, this (weakly convergent) expansion did not incorporate necessary boundary conditions, namely the correct value of $\varphi(\rho,z)$ outside the dot. This can be seen, e.g., from the facts that it yields a potential which falls off exponentially with $|z|$ instead of inversely, and that a plot of its $n(\rho)$ fails to properly approach Eq. (1) (see below) in the large-dot limit.

Apart from the analysis of impurity screening by an infinite sheet of electrons [27], to our knowledge the problem for a confined electron cloud has been correctly solved only in the large-dot limit $S \gg 1$, where $S = m^* e^{*2} R / \pi \hbar^2 = R / \pi a_B^*$. Here $a_B^*$ is the effective Bohr radius (we account for the semiconductor material's effective electron mass $m^*$, and for its dielectric constant by defining an effective charge $e^* = e/\varepsilon^{1/2}$). It has been shown [13] that this limit reflects a classical distribution of point charges confined to a potential well, and for the specific example of a harmonic confining potential $V = \gamma \rho^2$ (this is the most commonly assumed situation for a QD [28-31]; we set $V=0$ at the dot center) it leads to

$$n(\rho) = 4\gamma R / \pi^2 e^{*2} \sqrt{1 - \rho^2 / R^2} \tag{1}$$

This classical solution derives from the fact that a 2D charge distribution of the form (1) itself produces a quadratic in-plane potential [32], balancing the force from the above potential well *V*. This result was subsequently reproduced in Ref. [14] and used elsewhere, e.g. [33,34].

In the present Letter we outline the first full solution of the TF problem for a 2D radially confined quantum gas. This general solution is valid for an arbitrary potential $V(\rho)$ and for all values of the dot size parameter *S*. Note that while in many situations the parameter is relatively large, $S>1$ [13], this is not always the case. Importantly, it is shown below that there exists a wide parameter region ($0.1 \lesssim S \lesssim 5$) where perturbation theory with respect to *S* or $1/S$ is inapplicable.

**General solution.** - Assume that the electron density vanishes for $\rho>R$. It is convenient to write the TF equation in terms of the self-consistent electrostatic potential in the plane of the dot, $\varphi^*(\rho)$, defined via the effective (screened) charge:

$$e^*\varphi^*(x) = \tfrac{1}{2} S \int_0^1 dx_1 \hat{K}(x,x_1) \left[ \mu - e^*\varphi^*(x_1) - V(R\sqrt{x_1}) \right]. \tag{2}$$

Here $x \equiv \rho^2/R^2$ and $\hat{K}$ is the Coulomb interaction kernel, corresponding to $|\vec{\rho}-\vec{\rho}_1|^{-1}$. It can be expanded as follows:

$$\hat{K}(x,x_1) = \left(\frac{2}{x+x_1}\right)^{1/2} \sum_{j=0}^{\infty} \left(\frac{4xx_1}{(x+x_1)^2}\right)^j F(j), \tag{3}$$

$$F(j) = \frac{\Gamma(j+\tfrac{1}{4})\Gamma(j+\tfrac{3}{4})}{\Gamma^2(j+1)} \xrightarrow[j\gg 1]{} \frac{1}{j}\left(1 - \frac{3}{16j} - \frac{101}{1536j^2} + \ldots\right). \tag{4}$$

$F(j)$ is a slowly decreasing function. Applying the Euler-Maclaurin summation formula to its leading term one finds that the Coulomb kernel, Eq. (3), exhibits logarithmic divergence: its singular part equals $\ln[4/(x-x_1)^2]+\ldots$ .

In order to solve the integral equation (2) we define the eigenfunctions $\int_0^1 dx_1 \hat{K}(x,x_1)\psi_k(x_1) = \lambda_k \psi_k(x)$. The Kernel $\hat{K}(x,x_1)$ is symmetric, therefore the functions $\psi_{k=0,1,2,...}(x)$ may be taken to form an orthonormal basis: $\int_0^1 dx \psi_k(x)\psi_l(x) = \delta_{kl}$.

Near the dot center ($x \to 0$) the eigenfunctions can be expanded in a Taylor series in powers of $x^{1/2}$. Calculating the first odd coefficients explicitly, one finds that they vanish, suggesting that the expansion contains only integer powers of $x$. At the dot's outer edge, $\tilde{x} \equiv 1-x \ll 1$, there appears a singularity caused by the divergence of $\hat{K}(x,x_1)$ at $x=x_1$. In the main logarithmic approximation the singular parts can be summed, producing a modified Bessel function. Thus overall the eigenfunctions can be represented in the following form:

$$\psi_k(x) = W_k \left[ \phi_k^{(1)}(\tilde{x}) + \phi_k^{(2)}(\tilde{x}) I_0 \left( \sqrt{\frac{8}{\lambda_k} \tilde{x} ln\left(\frac{D}{\tilde{x}}\right)} \right) \right]. \tag{5}$$

Here $\phi_k^{(1)}$ and $\phi_k^{(2)}$ are functions regular in the regions $\tilde{x} \ll 1$ and $1-\tilde{x} \ll 1$, respectively, $D$ is a constant, and $W_k$ is the normalization factor. We make the assumption that $\phi_k^{(1)}$ and $\phi_k^{(2)}$ are polynomials of degree $k+1$ and $k$, respectively, with $\phi_k^{(1;2)}(\tilde{x}=0) = 0;1$. This is confirmed by the numerical determination of eigenfunction parameters (see the Appendix).

Employing the basis $\{\psi_k\}$, the solution of Eq. (2) is straightforward:

$$e^*\varphi^*(x) = \sum_{k=0}^{\infty} G_k \psi_k(x), \quad n(x) = \frac{m^*}{\pi\hbar^2}\left( \mu - V(R\sqrt{x}) - \sum_{k=0}^{\infty} G_k \psi_k(x) \right). \tag{6}$$

Here

$$G_k = (1-\sigma_k)\left[ \mu \int_0^1 dx \psi_k(x) - \int_0^1 dx \psi_k(x) V\left(R\sqrt{x}\right) \right], \quad \sigma_k \equiv \left(1 + \tfrac{1}{2} S \lambda_k\right)^{-1}. \tag{7}$$

The conditions that $N = \pi R^2 \int_0^1 n(x) dx$ and $n(x=1)=0$ set the chemical potential and the connection between $N$ and $R$. The solution for an arbitrary confining potential $V(\rho)$ is thus in principle fully defined. Instead of the second condition, which calls for calculating converged sums of weakly singular $\psi_k(x \to 1)$ terms, it is more convenient to employ integral forms of the eigenfunctions by imposing the condition that in the ground state $\partial E / \partial R = 0$. $E$ is the total energy (kinetic + electrostatic + confinement) of the electrons. For our 2D cloud satisfying the TF equation, it works out simply to

$$E = \frac{1}{2}\mu N + \frac{\pi}{2} R^2 \int_0^1 dx n(x) V(R\sqrt{x}). \tag{8}$$

**Harmonic confining potential.** - For the case of harmonic confinement $V = \gamma \rho^2$ one obtains

$$G_k = (1 - \sigma_k)(\mu \alpha_k - \gamma R^2 \beta_k), \tag{9}$$

where $\alpha_k \equiv \int_0^1 \psi_k(x) dx$ and $\beta_k \equiv \int_0^1 x \psi_k(x) dx$. Thanks to the completeness of the basis $\{\psi_k\}$, the latter satisfy useful exact relations:

$$\sum_{k=0}^{\infty} \alpha_k^2 = 1, \quad \sum_{k=0}^{\infty} \beta_k^2 = \tfrac{1}{3}, \quad \sum_{k=0}^{\infty} \alpha_k \beta_k = \tfrac{1}{2}, \quad \sum_{k=0}^{\infty} \alpha_k \psi_k(x) = 1, \quad \sum_{k=0}^{\infty} \beta_k \psi_k(x) = x. \tag{10}$$

Based on the above, a calculation finally leads to specific expressions for the dot radius, electron density distribution, and electron energy. It should be emphasized that apart from an overall prefactor, the following expressions have a universal character (the dot size enters only via the scaled quantities $S$ (or $\sigma_k$) and $x = \rho^2/R^2$):

$$N = \pi^4 \frac{\gamma}{e^{*2}} a_B^{*3} S^4 \zeta(S), \tag{11}$$

$$n(x) = \pi \frac{\gamma}{e^{*2}} a_B^* S^2 \left( \frac{z(S)}{\sum_{k=0}^{\infty} \sigma_k \alpha_k^2} \left( \sum_{k=0}^{\infty} \sigma_k \alpha_k \psi_k(x) \right) - \sum_{k=0}^{\infty} \sigma_k \beta_k \psi_k(x) \right), \tag{12}$$

$$E = \frac{\pi^6}{2} \frac{\gamma^2}{e^{*2}} a_B^{*5} S^6 \left( \frac{z^2(S)}{\sum_{k=0}^{\infty} \sigma_k \alpha_k^2} - \sum_{k=0}^{\infty} \sigma_k \beta_k^2 \right), \tag{13}$$

where

$$z(S) = \left( \sum_{k=0}^{\infty} \sigma_k \alpha_k^2 \right) \cdot \frac{2 \sum_{k=0}^{\infty} \sigma_k \alpha_k \beta_k - \tfrac{1}{4} S \sum_{k=0}^{\infty} \sigma_k^2 \alpha_k \beta_k \lambda_k}{\sum_{k=0}^{\infty} \sigma_k \alpha_k^2 - \tfrac{1}{4} S \sum_{k=0}^{\infty} \sigma_k^2 \alpha_k^2 \lambda_k}, \quad \zeta(S) = z(S) - \sum_{k=0}^{\infty} \sigma_k \alpha_k \beta_k. \tag{14}$$

Using the directly calculated values of $\{\psi_k, \lambda_k\}_{k=0\text{-}4}$, accurate extrapolation formulae can be developed for $\alpha_k$, $\beta_k$, $\lambda_k$ for convenient use in the above equations (see the Appendix). Fig. 1 illustrates electron density distributions within a 2D parabolic quantum dot. As befits the semiclassical treatment, the shapes match the spatial average of a numerical solution of the 2D Schrödinger equation for the same system [35]; the electron cloud radii are in excellent agreement. Fig. 2 shows the variation of the dot radius as a function of its electron number. Fig. 3(a) depicts the evolution of the total internal energy of the confined electrons. Hartree-Fock and diffusion Monte Carlo values from Ref. [36] are superimposed within their available range, demonstrating very good agreement.

**Application: Capacitive energies.** - Coulomb blockade experiments summarized in Ref. [37] measured the so-called "capacitive" (or "addition") energy for electrons confined to circular quantum dots within semiconductor pillar structures. This quantity can be represented [38] as the second difference $E_{cap}=E(N+1)+E(N-1)-2E(N)$ which can be directly calculated from the internal energy expression (13). Fig.3(b) shows that the analytical calculation yields an excellent match with the data [37].

**Limiting cases.** - The full solution above can be simplified in the limiting cases of small and large values of the *S* parameter (½$S\lambda_k$<<1 or >>1, equivalent to $\sigma_k$→1and 0 respectively), when the various series can be written out in powers of *S* or 1/*S*. The limiting cases correspond to *S*<<0.1 and *S*>>5.[†] Thus, for example, the electron density (12) can be written generally as $n(x) = (\gamma R / e^{*2}) Z(x, S)$. The large-*S* limit is $Z = (4/\pi^2)(1-x)^{1/2}$, according to (1) and independent of *S*, while for very small *S* we find *Z*=*S*(1-*x*). Analogously, from Eq. (11) we find two very different relations between the number of electrons in the dot and its radius: $N_{S>>5} = (8\gamma / 3\pi e^{*2}) R^3$ and $N_{S<<0.1} = (m^*\gamma / 2\hbar^2) R^4$.

As mentioned in the Introduction, it is perfectly realistic for *S* to lie below the classical solution region. Systems illustrated in Figs. 1 and 3 offer examples. The lower bound of *S* follows from the TF condition *N*>>1. Using Eq. (11) and noticing that $\zeta(S)$ decreases slowly from $\zeta(0)$=0.5, this leads to $S^4 >> 2e^{*2} / (\pi^4 \gamma a_B^{*3})$.

**Conclusion.** - We have presented a general and consistent solution of the semiclassical TF equation for quantum dot "artificial atoms" describing a 2D (transversely restricted) electron gas cloud confined by a radial potential *V*(*ρ*). The analytical solution accounts both for the specifically 2D density of states of the electrons and for the 3D Poisson equation satisfied by their electrostatic potential. Its mathematical structure is peculiar, since the in-plane Coulomb kernel and its eigenfunctions exhibit logarithmic singularities at the dot boundary. This requires care in formulating the equations for the dot radius, electron density, energy, and other system parameters. As an example, equations are presented for the common situation of a harmonic $V(\rho)$ [Eqs. (11)-(13)]. Furthermore, it turns out that there is a wide range of dot size parameters $S \equiv R/\pi a_B^*$ where neither a large- nor a small-dot approximation is valid. In fact, the classical limit where the electrons' electrostatic energy dominates over the kinetic energy of quantum degeneracy is not reached until $S \gtrsim 5$.

---

[†] The small-*S* range is determined by the fact that $\lambda_0$≈11. In the opposite limit, the leading 1/*S* terms have coefficients of the form $\sum_k (\alpha_k^2 / \lambda_k^m)$, $\sum_k (\beta_k^2 / \lambda_k^m)$, $\sum_k (\alpha_k \beta_k / \lambda_k^m)$. For *m*=1,2 these can be calculated from the numbers and extrapolations in the Appendix, while an estimate of the remainder yields ~$S^{-3}\ln S$. The large-*S* range is attained when this is <<1.

The solution developed here can be applied to arbitrary confining potentials $V(\rho)$. Furthermore, it can be extended to describe "breathing mode" oscillations of the electron cloud. These topics will be addressed in a forthcoming publication. The general approach may also have utility for electron islands at liquid helium surfaces [40] and on graphene [41], ions in "pancake" traps [42], and π-electrons in conjugated molecules [43].

***

We appreciate very useful advice by A. V. Chaplik, S. Nazin, and V. Shikin, and discussions with H. Saleur and J. Yngvason. This work is supported by the NSF (A.H. and V.V.K., grant PHY-1068292,) and by EOARD (Yu.O., grant 097006).

APPENDIX

*Eigenfunctions.* - Eigenfunctions of the form given in Eq. (5) were determined by varying the coefficients of the polynomials comprising $\phi_k^{(1)}$ and $\phi_k^{(2)}$, as well as the quantity $D$ and the eigenvalues $\lambda_k$ so as to minimize the functional

$$\Theta_k = \frac{\int_0^1 dx \left( \psi_k(x) - \frac{1}{\lambda_k} \int_0^1 dx_1 \hat{K}(x,x_1) \psi_k(x_1) \right)^2}{\int_0^1 dx \psi_k^2(x)} . \tag{A.1}$$

This was done successively for $k=0,1,2,\ldots$ while keeping the eigenfunction $\psi_k$ orthogonal to the subspace $\{\psi_0, \ldots \psi_{k-1}\}$. For this we implement the constraints $\int_0^1 \psi_k(x)\psi_l(x)dx = \delta_{kl}$ for all $l=0,\ldots,k-1$. In solving for $\psi_k$ these conditions reduce the number of free parameters by $k$. We evaluated the first five ($k=0$-4) orthonormalized eigenfunctions of the Coulomb kernel $\hat{K}(x,x_1)$. In the following list we provide the values of the coefficients of the polynomials $\phi_k^{(1)} = \sum_{n=1}^{k+1} a_n \tilde{x}^n$ and $\phi_k^{(2)} = \sum_{n=0}^{k} b_n \tilde{x}^n$, the eigenvalues $\lambda_k$, the normalization constants $W_k$, the integral quantities $\{\alpha_k, \beta_k\}$, and the parameter $\Theta_k$ whose smallness characterizes the closeness of the eigenfunctions to the actual basis: $\{(a_1,\ldots,a_{k+1}),(b_0,\ldots,b_k), W_k, \lambda_k, \alpha_k, \beta_k, \Theta_k\}_k$ = $\{(0.7), (1), 0.69, 10.82, 0.989, 0.451, 6\times10^{-5}\}_{k=0}$; $\{(-9.83, -5.79), (1, 8.04), 0.75, 2.88, 0.126, 0.314, 4\times10^{-4}\}_{k=1}$; $\{(-9.02, 89.0, -3.5), (1, 2.5, -47.8), 0.85, 1.65, 0.061, 0.137, 2\times10^{-3}\}_{k=2}$; $\{(39.31, 533.25, -530.70, 32.9), (1, -48.1, -138.1, 145.8), 0.91, 1.15, 0.038, 0.081, 7\times10^{-3}\}_{k=3}$; $\{(340.31, 2189.3, -9147.42, 4294.51, 206), (1, -325, -393, 1038, -218), 0.90, 0.87, 0.026, 0.053, 1\times10^{-2}\}_{k=4}$. The insert in Fig. 1 shows plots of these eigenfunctions.

*Extrapolation formulae for the quantities $\lambda_k$, $\alpha_k$, $\beta_k$.* - The preceding section lists the numerically obtained values of the first five sets of eigenfunctions parameters. We can also establish accurate extrapolation formulae for $\lambda_{k>4}$. Start by noting that by virtue of Eq. (3)

$$\text{Tr}(\hat{K}) = \int_0^1 \hat{K}(x,x)dx = 2\sum_{j=0}^{\infty} F(j) . \tag{A.2}$$

From Eq. (4), $F(j\gg1)\to j^{-1}$ and $\left[\mathrm{Tr}(\hat{K})\right]_N \equiv 2\sum_{j=0}^{N} F(j) = 2\ln N + 9.4722$. Since also $\left[\mathrm{Tr}(\hat{K})\right]_N \equiv \sum_{k=0}^{N} \lambda_k$, the expected behavior of the eigenvalues is $\lambda_{k\gg1}\to 2/k$ and we seek an approximation of the form

$$\lambda_k = \frac{2}{k+G_\lambda}\left(1 + \frac{1}{C_\lambda^{(0)} + C_\lambda^{(1)}k + C_\lambda^{(2)}k^2}\right). \tag{A.3}$$

(There is a weak odd-even variation in the first five numerical values of $\lambda_k$, but it may be neglected in the above formula.) The coefficients can be optimized based on the set of $\lambda_{k=0\text{-}4}$ values listed above, on the numerically computed magnitude of

$$\sum_{k=0}^{\infty} \lambda_k^2 = \mathrm{Tr}(\hat{K}^2) = \int_0^1 dx \int_0^1 dx_1 \hat{K}^2(x,x_1) = 134.628, \tag{A.4}$$

and on a comparison of the expansions of $E$, $\mu$ and $N(R)$ to first order in $1/S$ with the classical limit following from Eq. (1). One finds $G_\lambda = -0.99$, $C_\lambda^{(0)} = 24$, $C_\lambda^{(1)} = -7.865$, $C_\lambda^{(2)} = 0.668$.

Analogous formulae can be set up for the sums $\alpha_k$ and $\beta_k$ for $k>4$. By using the first three sum rules in Eq. (10) and by again comparing the expansions of $E$, $\mu$ and $N(R)$ with the classical limit, one can deduce the constraint $\lim_{k\to\infty}(\beta_k/\alpha_k) = 2$ and the following extrapolation parameters:

$$\alpha_k \approx \frac{0.741}{k^2 + 2.236k + 3.842}, \quad \beta_k \approx \alpha_k\left(2 - \frac{0.0455}{k - 4.717}\right). \tag{A.5}$$

The accuracy of the above approximations can be estimated by combining them with the numerically calculated values listed above for $0\le k\le 4$ and comparing the resulting sums with their exact values in Eq. (10). The differences are found to be satisfyingly small: $4\times 10^{-6}$ for $\sum \alpha_k^2$, $1\times 10^{-4}$ for $\sum \alpha_k \beta_k$ and $3\times 10^{-4}$ for $\sum \beta_k^2$.

*Electron density plots.* - When calculating integrals of the density, using the series (12) produces very fast convergence. For example, in the large-$S$ limit even if the sum is restricted just to the five terms with numerically evaluated parameters, the remainder comprises only $10^{-4}$ of the entire sum. However, convergence of the series for the density itself, $n(x)$ for some fixed value of $x$ (particularly near $x=1$), is rather slow. To improve the convergence of the series, one can make use of the last two identities in Eq. (10) by performing an addition and subtraction:

$$\tilde{n}(x) = n(x) + \frac{\gamma R}{e^{*2}} \left[ Q_1 \left( 1 - \sum_{k=0}^{\infty} \alpha_k \psi_k(x) \right) + Q_2 \left( x - \sum_{k=0}^{\infty} \beta_k \psi_k(x) \right) \right], \quad (A.6)$$

where $n(x)$ is the solution (12). If all the eigenvalues and eigenfunctions were known and summed exactly, the extra terms would make zero contribution. But in our case only the first five sets plus the extrapolation formulae are available, and the optimal coefficients $Q$ can be found by breaking up the sums involving $\psi_k$ in Eqs. (12) and (A.6) into two parts: $\tilde{n}_1(x)$ for $k$=0-4 and $\tilde{n}_2(x)$ for $k \geq 5$. By imposing the requirements that the former part vanish at $x$=1 and the norm of the latter part be minimized, $Q_1$ and $Q_2$ can be determined. For large $S$, a calculation yields $Q_1$=-11.406 and $Q_2$=5.544. The resulting density profile, $\tilde{n}_1(x)$, is the one plotted in Fig. 1.

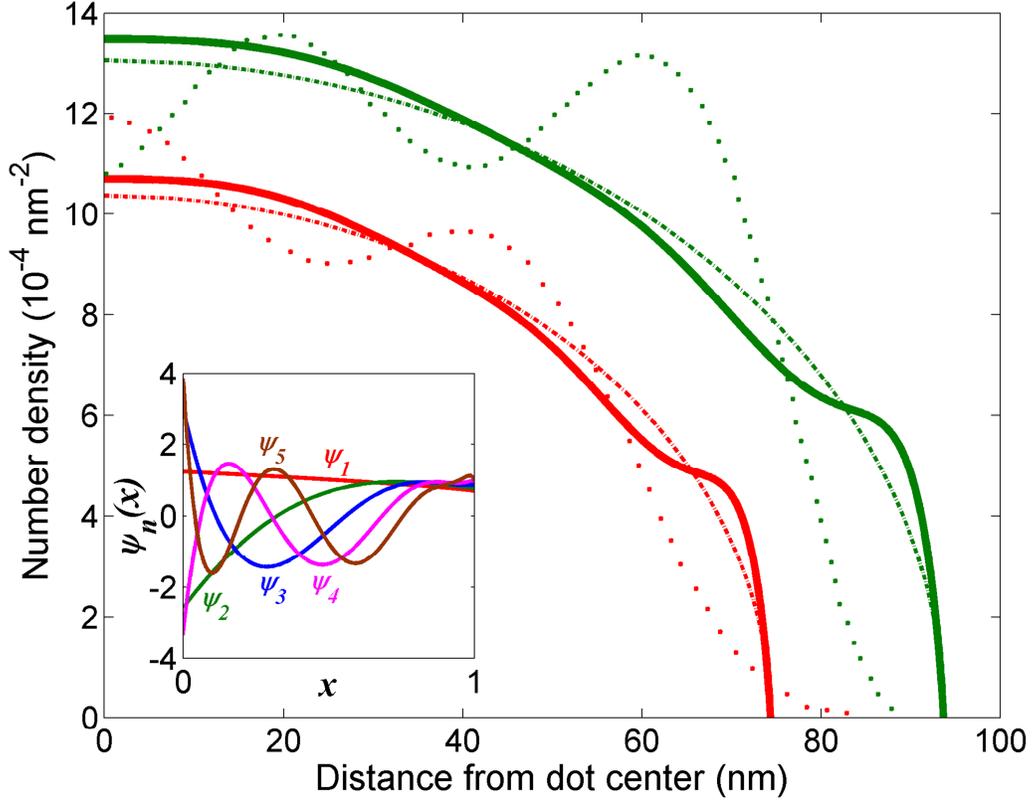

**Fig. 1.** Electron density distributions for a 2D parabolic-confinement quantum dot in In$_{0.05}$Ga$_{0.95}$As ($m^*$=0.0648$m_e$, $\varepsilon$=12.98, $\gamma$=3.82 µeV/nm$^2$) for $N$=12 (red curves, bottom, corresponding to $S$=2.23) and $N$=24 electrons (green curves, top, $S$=2.81). The dimensionless size parameter $S = m^* e^{*2} R / \pi \hbar^2 = R / \pi a_B^*$. Solid line: solution of the Thomas-Fermi equation based on the first five terms, $k$=0-4, of the series in Eq. (12) (see also the Appendix); dashed line: Eq. (1); dotted line: numerical solution of the Schrödinger equation [35]. It should be emphasized that while the shape of the density function will evolve slightly if higher-$k$ terms are added to the series in a straightforward manner, the radius of the QD and the quantities involving integrals of the density are already given with very high accuracy by just the first few terms. Inset shows the orthonormalized eigenfunctions, Eq. (5).

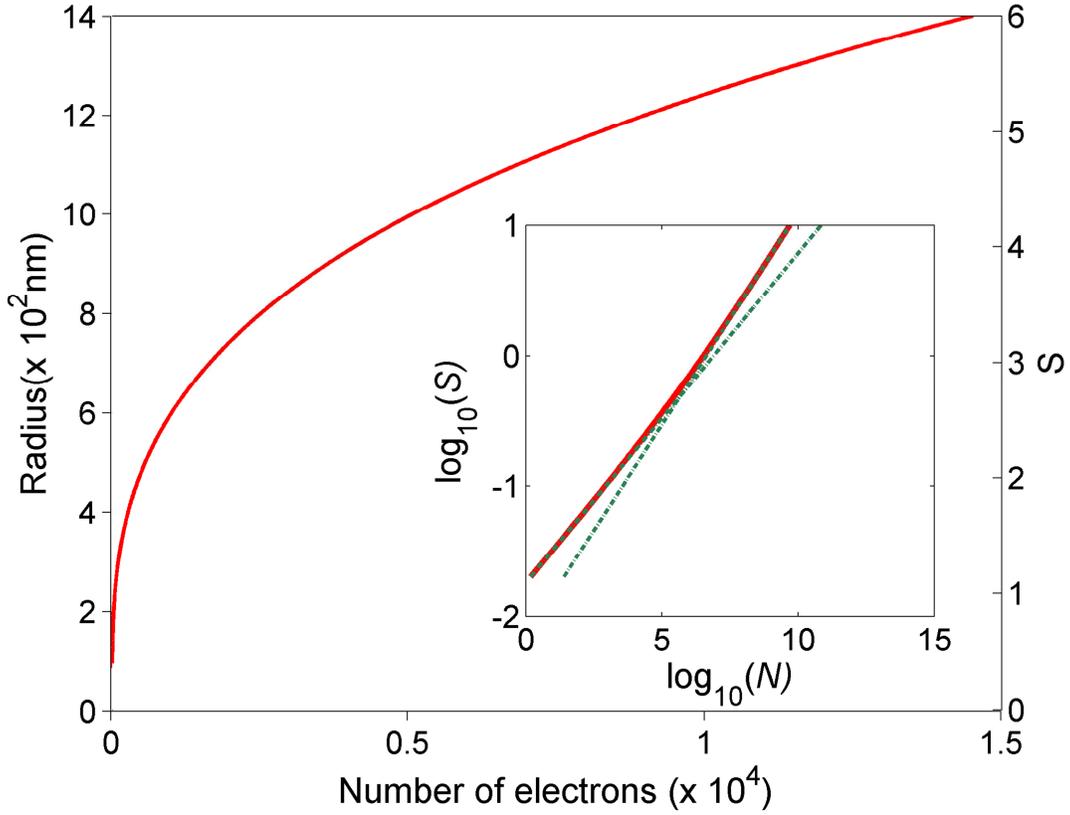

**Fig. 2.** Radius of the 2D electron cloud in InSb semiconductor ($m^*=0.013m_e$, $\varepsilon=16.8$, $\gamma=0.77$ μeV/nm$^2$) as a function of the number of electrons filling the parabolic quantum dot, as given by Eq. (11). Inset: Plot of $\log_{10}(S)$ vs. $\log_{10}(N)$ over a wider range, corresponding to $0.02<S<10$ (here $\gamma$ was set to 55 meV/nm$^2$ in order to cover this $S$ range). As indicated by the dashed lines, the dependence $N \propto R^t$ changes from $t=3.95$ to $t=3.06$ with increasing dot size, in agreement with the asymptotic limiting values of 4 and 3 as described in the text.

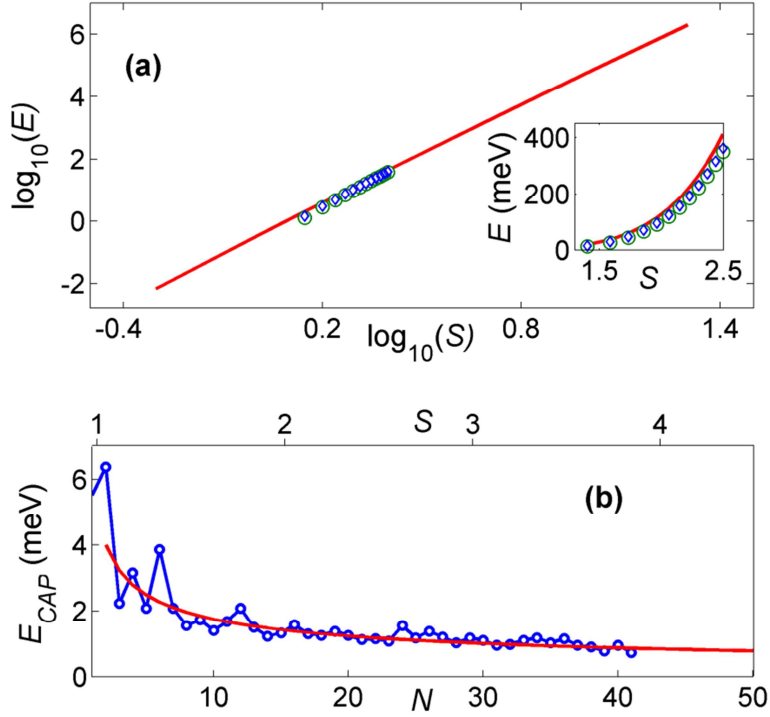

**Fig. 3.** (a) Total energy of electrons confined to a circular parabolic quantum dot, Eq. (13) with $E$ in units of $\pi^6 \gamma^2 a_B^{*5}/2e^{*2}$ [the prefactor of Eq. (13).]. Numerical results for GaAs dots ($\varepsilon=12.4$, $m^*=0.067m_e$, $\gamma=4.99$ μeV/nm$^2$, $N=2$-13) [36] are shown for comparison as circles (Monte Carlo) and diamonds (Hartree-Fock). This region is enlarged in the inset. In the limit of $S \gg 5$ the energy becomes purely electrostatic, but in the highlighted region both kinetic and potential energy contributions need to be considered and perturbation theory with respect to $1/S$ does not hold. (b) Capacitive (addition) energy for electrons in an In$_{0.05}$Ga$_{0.95}$As ($m^*=0.0648m_e$, $\varepsilon=12.98$) quantum dot. Dots: experimental data [37]. Line: values obtained from the semiclassical energy expression in the text. The dominant contribution to the energy in the present case comes from the electrostatic terms. The curvature of the confining potential is [39] $\gamma = (e^*)^2/(2r_0^3 N^{1/2})$ with $r_0 = 1.5 a_B^*$. (The peaks in the data at $N=2,6,12$ are due to quantum 2D shell closings.)